\documentclass[epsf,graphics,a4,12pt]{article}
\usepackage{epsfig}
\begin{document}
\baselineskip=18pt
\newcommand{\be}{\begin{equation}}
\newcommand{\en}{\end{equation}}
\begin{center}
{\large Black Hole entropy by the brick wall method  in four and five
  dimensions with $U(1)$ charges} 
\end{center}
\vspace{1ex}
\centerline{\large
Elcio Abdalla and L. Alejandro Correa-Borbonet}
\begin{center}
{Instituto de F\'{\i}sica, Universidade de S\~{a}o Paulo,\\
C.P.66.318, CEP 05315-970, S\~{a}o Paulo, Brazil}
\end{center}
\vspace{6ex}
\begin{abstract}
Using the brick wall method we compute the statistical entropy of a scalar 
field in a nontrivial background , in two different cases. These
background are generated by four and five dimensional black holes with
four and three $U(1)$ charges respectively. The Bekenstein entropy
formula is generally obeyed, but corrections are discussed in the
latter case. 
\end{abstract}
\vspace{6ex} \hspace*{0mm} PACS number(s): 04.70.-s;04.70.Dy;04.50.+h;04.62.+v.
\vfill

In a seminal paper$\cite{beken}$ Bekenstein proposed, several years
ago, that the entropy in black holes is proportional to the surface
area of the event horizon. Such an idea, motivated by properties of
the black hole area, which  
are very similar to the laws of thermodynamics, generated a whole
field of research. It has been proved that the proportionality between
entropy and area in black holes has far reaching consequences as the
generalized second law(GSL) of thermodynamics $\cite{brus}$. 
Moreover, in the context of quantum field theory Hawking
found$\cite{hawk}$ that black holes generate also  a thermal
radiation, and the related entropy is also given, in such a
derivation, in terms of the area of the event horizon, that is
$S=A/4G$, where $G$ is the Newton's constant.

Furthemore, recently, 't Hooft$\cite{thoo}$ and Susskind$\cite{lsuss}$
proposed the powerful holographic principle, which generalizes the
previous set up to general situations including quantum gravity,
foreseeing that entropy is  
always bounded by the surface area, that is $S\leq A/4G$. Such a 
principle is a truly new way of thinking quantum gravity, and sets up a road 
to quantization. It has also been considered in other physical situations, 
such as cosmological models$\cite{sussk}$. We should also mention that this 
idea has received an strong support from the conjeture known as the
AdS/CFT correspondence$\cite{malda}$. 

Althought our  knowledge about the black hole entropy, from the
thermodynamical point of view, is established, we do not have a
complete answer to  
this question in the statistical mechanics framework. A very well
known approach to this question was introduced by 't
Hooft$\cite{hooft}$. He showed how the entropy can be obtained from a
thermal bath of fields propagating just outside the horizon. This 
is the socalled brick wall method.

The purpose of this paper is to calculate the entropy of a scalar field 
propagating in backgrounds obtained in the string theory context$\cite{bhole}$.

We should also stress that although the degrees of freedom for black
holes in the string theory are explicitly known $\cite{vafa}$ we wish
to check the validity of the  't Hooft approach for these new
metrics. 

Initialy we consider the $D=4$ black hole with four $U(1)$
charges. This can be obtained upon toroidal compactification  
from two different intersecting M-branes configurations$\cite{kle}$, 
\be
ds^{2}=-\lambda^{-1/2}hdt^{2}+\lambda^{1/2}[h^{-1}dr^{2}+r^{2}
(d\theta^{2}+\sin ^{2}\theta d\varphi^{2})],
\label{eq:metri}
\en
where the $h(r)$-function describes the event horizon, similarly to
the Schwarzs\-child black hole, that is, 
\be
h(r)=1-\frac{r_{0}}{r},
\en
while the new $\lambda(r)$-function is characterized by the four
$U(1)$ charges, being given by  
the expression
\be
\lambda(r)=(1+\frac{r_{0}sinh^{2}\alpha_{2}}{r})(1+\frac{r_{0}sinh^{2}
\alpha_{5}}{r})(1+\frac{r_{0}sinh^{2}
\alpha_{6}}{r})(1+\frac{r_{0}sinh^{2}\alpha_{p}}{r}).
\en
In this background, we consider a  minimally coupled scalar field
which satisfies the  
Klein-Gordon equation$\cite{birrell}$
\be
\frac{1}{\sqrt{-g}}\partial_{\mu}(\sqrt{-g}g^{\mu \nu}
\partial_{\nu}\Phi)-m^{2}\Phi=0 .\label{eq:gordon}
\en
The 't Hooft method consists in introducing a brick wall cut-off near 
the event horizon such that the boundary condition
\be
\Phi=0 \;\;\;\;\; for   \;\;\;\;\; r\leq r_{0}+\epsilon
\en
is satisfied.
In order to eliminate infrared divergences another cut-off is introduced at
a large distance from the horizon, $L\gg r_{0}$, where we have,
\be
\Phi=0 \;\;\;\;\; for \;\;\;\;\; r\geq L.
\en 
In the spherically symmetric space, the scalar field can be decomposed as
\be
\Phi(t,r,\theta,\varphi)=e^{-iEt}R(r)Y(\theta,\varphi).
\en
Substituting this expresion back into (\ref{eq:gordon}) we obtain,
after some manipulations, the radial equation  
\be
\lambda^{1/2}h^{-1}E^{2}R(r)+\frac{1}{\lambda^{1/2}
r^{2}}\partial_{r}[r^{2}h\partial_{r}R(r)]-\frac{l(l+1)}
{\lambda^{1/2}r^{2}}R(r)-m^{2}R(r)=0.\label{eq:radial}
\en
Using the WKB approximation, we consider $R(r)=\rho(r) e^{iS(r)}$, 
the function $\rho(r)$ is a slowly varying amplitude and $S(r)$ is a
rapidly varying phase. To leading order, only the contribution from
the first derivatives of $S$ are important.  
Then, from eq($\ref{eq:radial}$) we get the radial wave number
$K\equiv \partial_{r}S$: 
\be
K=\lambda^{1/2}\left ( 1-\frac{r_{0}}{r} \right)^{-1}\sqrt{E^{2}-
\left( 1-\frac{r_{0}}{r}\right)\left(
\frac{l(l+1)}{\lambda r^{2}}+\frac{1}{\lambda^{1/2}}m^{2}\right)}.
\en
In such a case, then the number of radial modes $n_{r}$ is given by
\be
\pi n_{r}=\int^{L}_{r_{0}+\epsilon } dr K(r,l,E),\label{eq:estados}
\en
In order to find the entropy of this system, we calculate the 
free energy of a thermal bath of scalar particles with an inverse temperature 
$\beta$, that is
\be
e^{-\beta F}=\sum e^{-\beta E_{N_{\tau}}}
\en
where $E_{N_{\tau}}$ is the total energy corresponding to the 
quantum state $\tau$. The sum also includes the degeneracies of the
quantum states, therefore we have   
\be
e^{-\beta F}=\prod _{(n_{r},l,m_{l})} \frac{1}{1-\exp (-\beta E)}.
\en
where $(n_{r},l,m_{l})$ are the set of quantum numbers associated to this 
problem. The product $\prod $ take into account the contribution from all the
modes. The factor $(1-e^{-\beta E})^{-1}$ is due to the fact that we
are dealing with bosons where the ocupation number can take on the
value of all positive integers as well as zero, so that 
\be
\sum^{\infty}_{n=0} e^{-\beta nE}=\frac{1}{1-\exp(-\beta E)}
\en 
From the previous equation we can write the free  energy as
\begin{eqnarray}
 F & = &
\frac{1}{\beta} \sum_{n_{r},l,m_{l}} log(1-e^{-\beta E}) \nonumber\\
 & = & 
  \frac{1}{\beta}\int dl (2l+1)\int dn_{r}log(1-e^{-\beta E}).
\end{eqnarray}
Integrating by parts and using (\ref{eq:estados}) we get
\begin{eqnarray} 
 F & = & 
 -\int dl (2l+1) \int dE \frac{1}{exp(\beta E)-1} n_{r} \nonumber\\
   & = &
 -\frac{1}{\pi}\int dl(2l+1)\int dE 
\frac{1}{\exp (\beta E)-1} \int^{L}_{r_{0}+\epsilon} dr
 \lambda^{1/2}(r)\nonumber \\ & & \times 
\left( 1-\frac {r_{0}}{r} \right)^{-1}\sqrt{E^{2}-\left( 1-\frac
 {r_{0}}{r} \right)\left( \frac{l(l+1)}{\lambda
 r^{2}}+\frac{m^{2}}{\lambda^{1/2}}\right)}\label{eq:esta} 
\end{eqnarray}
The $l$ integration can be perfomed explicitly and it is taken only
over those values for which the square root exits, 
\begin{eqnarray} 
\lefteqn{\int dl (2l+1) \left[ E^{2} -\left( 1-\frac {r_{0}}{r} \right)
\left(\frac {l(l+1)}{\lambda r^{2}} +\frac{m^{2}}{\lambda^{1/2}}\right)
\right] ^{\frac{1}{2}} =} \nonumber \\
 & & \frac  {2}{3} \lambda r^{2} \left( 1-\frac {r_{0}}{r} \right) ^{-1}
\left( E^{2} -\left( 1-\frac {r_{0}}{r} \right) \frac{m^{2}}{\lambda^{1/2}}
\right) ^{\frac {3}{2}};
\end{eqnarray}
therefore, we get for the free energy the result
\be
F=-\frac{2}{3\pi}\int dE \frac{1}{\exp (\beta E)-1}\int^{L}_{r_{0}+\epsilon} dr
r^{2}\lambda^{3/2}\left(
  1-\frac{r_{0}}{r}\right)^{-2}\left[E^{2}-\left( 1-\frac{r_{0}}{r} 
\right)\frac{m^{2}}{\lambda^{1/2}} \right]^{3/2}.\label{eq:free}
\en
We can see that the integrand diverges at the event horizon. Using the brick 
wall method we shall give a meaning to such a divergence as a renormalization 
of the Newton's constant. We first introduce the variable,
$y=r/r_{0}$. Substituting it back into (\ref{eq:free}) we find, after
some trivial substitutions,   
\be
F=-\frac{2r^{3}_{0}}{3\pi}\int dE \frac{1}{\exp(\beta E)-1} 
\int^{\bar{L}}_{1+\bar{\epsilon}} dy y^{2}
\lambda^{3/2}(y) \left( 1-\frac{1}{y}\right)^{-2}\left[E^{2}-
\left( 1-\frac{1}{y}\right)\frac{m^{2}}{\lambda^{1/2}(y)}
\right]^{3/2}\label{eq:diezseis}, 
\en
where $\bar{\epsilon}=\frac{\epsilon}{r_{0}},\bar{L}=\frac{L}{r_{0}}$.

The contribution of large values for $y$ to the previous integral is
\be
F_{L}=-\frac{2L^{3}}{9\pi}\int^{\infty}_{m}dE \frac{(E^{2}-
m^{2})^{3/2}}{\exp(\beta E)-1}.
\en
We can recognize that this is the usual term arising  from the vacuum
surrounding the 
system at large distances.

Near the horizon, that is for $y$ near $1$, we find the expression  
\be
F_{\epsilon}= -\frac{2r^{3}_{0}\lambda^{3/2} (1)}{3\pi}\int^{\infty}_{0} dE 
\frac{E^{3}}{\exp(\beta E)-1}
\int^{\bar{L}}_{1+\bar{\epsilon}} dy (y-1)^{-2}.
\en
We next use the well known formula
\be
\int^{\infty}_{0} dE \frac{E^{3}}{e^{\beta E}-1}=\frac{\pi^{4}}{15\beta^{4}}
\label{eq:energia},
\en
and integrate over $y$. The expression for $F_{\epsilon}$ reduces to
\be
F_{\epsilon}=-\frac{2\pi^{3}}{45\epsilon} 
\frac{\lambda^{3/2}(r_{0})r^{4}_{0}}{\beta^{4}}
\en
allowing us to directly compute the entropy from
\be
S_{\epsilon}=\beta^{2}\frac{\partial F_{\epsilon}}{\partial
  \beta}=\frac{8\pi^{3}}{45\epsilon} 
\frac{\lambda^{3/2}(r_{0})r^{4}_{0}}{\beta^{3}}.
\en
We need only to obtain  the inverse of temperature, that is $\beta$.
The Hawking temperature of the black hole is defined as
\be
T_{H}=\beta^{-1}_{H}=\frac{\kappa_{s}}{2\pi},
\en
where $\kappa_{s}$ is the surface gravity, given in terms of the
metric by the expression  
\be
\kappa_{s}=\left[\sqrt{g^{rr}}(\partial_{r}\sqrt{-g_{tt}})\right]_{r=r_{0}}.
\en
For the metric$(\ref{eq:metri})$ the surface gravity is thus
\be
\kappa_{s}=\frac{1}{2r_{0}\lambda^{1/2}(r_{0})}.
\en
Therefore, the Hawking temperature is obtained, from which we
subsequently find the entropy, that is, 
\be
\beta_{H}=4\pi r_{0}\lambda^{1/2}(r_{0}),
\en
and
\be
S=\frac{r_{0}}{360\epsilon}.
\en
This expression can be transformed making use of the invariant distance
\be
\int ds =\int^{r_{0}+\epsilon}_{r_{0}}
dr\frac{\lambda^{1/4}(r)}{\sqrt{1-r_{0}/r}}= 
\sqrt{4\epsilon r_{0}\lambda^{1/2}(r_{0})},
\en
in terms of which we can rewrite the entropy over area relation as a function 
of invariants only, 
\be
S=\frac{r_{0}}{360\epsilon}=\frac{A}
{1440\pi \epsilon r_{0}\lambda^{1/2}(r_{0})}=\frac{A}{360\pi(\int ds)^{2}},
\en
where $A=4\pi r^{2}_{0}\lambda^{1/2}(r_{0})$ is the horizon area.

Therefore the entropy of the scalar field is proportional to the area
and diverges linearly with the cutoff $\epsilon$ (or quadratically with
the invariant distance). The nature of this divergence can be
understood in terms of the infinite gravitational redshift between the
horizon and infinity. Any field mode with finite frequency at the
horizon must have vanishing frequency at infinity. Therefore, the
number of states with arbitrarily small energy is infinity. The
regulator $\epsilon$ is an ultraviolet regulator, which has the
function of cutting off the sum over states.   

In reference $\cite{uglum}$ it was showed that the question of
finiteness of the entropy can be solved by the renormalization of
Newton's gravitational constant. We thus can obtain the
Bekenstein-Hawking entropy formula, 
\be
S_{bare}+S_{\epsilon}=(\frac{1}{4G_{N}}+\frac{1}{360\pi (\int
  ds)^{2}})A=\frac{A}{4G_{R}}. 
\en

We would like to know whether these results could be obtained in other
situations. 
That is the reason why we also consider the case of a minimally
coupled scalar field in the background of a $D=5$ black hole with
three $U(1)$ charges. This black hole can be obtained from a
configuration with the intersection of D-$5$ branes and D-$1$
branes$\cite{horow}$. Obviously the calculation for the black hole in
five dimensions has many features in common with the $D=4$ black hole
considered above. In this case the metric reads 

\be
ds^{2}_{5}=-f^{-2/3}hdt^{2}+f^{1/3}[h^{-1}dr^{2}+r^{2}d\Omega^{2}_{3}]
\label{eq:cinco}
\en
where
\be
h(r)=1-\frac{r^{2}_{0}}{r^{2}}
\en
and the $U(1)$ charges are represented in the function $f(r)$ as
\be
f(r)=(1+\frac{r^{2}_{0}sinh^{2}\alpha
  }{r^{2}})(1+\frac{r^{2}_{0}sinh^{2}\gamma }{r^{2}}) 
(1+\frac{r^{2}_{0}sinh^{2}\sigma }{r^{2}}).
\en
Once more we use the Ansatz $\Phi(t,r,\theta,\psi, \zeta)=e^{-iEt}R(r)Y(\theta,\psi,\zeta)$ 
in order to solve the $5$ dimensional generalization of (\ref{eq:gordon}), a 
procedure which leads to the radial equation
\be
f^{2/3}h^{-1}E^{2}R(r)+\frac{1}{f^{1/3}r^{3}}
\partial_{r}[r^{3}h\partial_{r}R(r)]-
\frac{l(l+2)}{f^{1/3}r^{2}}R(r)-m^{2}R(r)=0,
\en
with the corresponding radial wave number
\be
K=f^{1/2}\left ( 1-\frac{r^{2}_{0}}{r^{2}} \right)^{-1}
\sqrt{E^{2}-\left( 1-\frac{r^{2}_{0}}{r^{2}}\right)\left(
\frac{l(l+2)}{f r^{2}}+\frac{1}{f^{2/3}}m^{2}\right)}
\en
Therefore, following the same vein as described in
($\ref{eq:estados}$) through  
($\ref{eq:esta}$) we find for the free energy the integral 
\begin{eqnarray}
 F & = &
\frac{1}{\beta} \int dl(l+1)^{2}\int dn_{r} log(1-e^{-\beta E}) \nonumber\\
   & = & -\frac{1}{\pi}\int dl(l+1)^{2}\int dE 
\frac{1}{\exp (\beta E)-1} \int^{L}_{r_{0}+\epsilon} dr
f^{1/2}(r)\nonumber \\ & & \times 
\left( 1-\frac {r^{2}_{0}}{r^{2}} \right)^{-1}\sqrt{E^{2}-\left(
    1-\frac {r^{2}_{0}}{r^{2}}  
\right)\left( \frac{l(l+2)}{f r^{2}}+\frac{m^{2}}{f^{2/3}}\right)}
\end{eqnarray}
As before, we can perform the $l$ integration explicitly, obtaining
three terms, that  
is,
\begin{eqnarray}
\lefteqn{\int dl (l+1)^{2}\sqrt{E^{2}-\frac{hl(l+2)}{fr^{2}}-
\frac{hm^{2}}{f^{2/3}}} = } \nonumber\\
  & & \frac{f^{3/2}r^{3}(E^{2}-\frac{hm^{2}}{f^{2/3}}+\frac{h}
{fr^{2}})^{2}}{8h^{3/2}}
[\frac{\pi}{2}-arcsin(\sqrt{\frac{h/fr^{2}}{E^{2}-\frac{hm^{2}}{f^{2/3}}+
\frac{h}{fr^{2}}}})]+ \nonumber \\
  & & +\frac{fr^{2}(E^{2}-\frac{hm^{2}}{f^{2/3}})^{3/2}}{4h}-\frac{fr^{2}(E^{2}
  -\frac{hm^{2}}{f^{2/3}}+\frac{h}{fr^{2}})\sqrt{E^{2}-
\frac{hm^{2}}{f^{2/3}}}}{8h} .
\end{eqnarray}
%
First let us calculate the contribution to the free energy, near the
horizon, from 
the first term of the previous equation, that is,
\be
F^{(5)}_{\epsilon}=-\frac{r^{4}_{0}f^{2}(r_{0})}{32}\int^{\infty}_{0}
dE \frac{E^{4}}{exp(\beta E)-1} 
\int^{\bar{L}^{2}}_{(1+\bar{\epsilon})^{2}} dy (y-1)^{-5/2}
\en
where $y=r^{2}/r^{2}_{0}$.

Now we use the formula
\be
\int^{\infty}_{0} dE \frac{E^{4}}{exp (\beta E)-1}=\frac{24\zeta
  (5)}{\beta^{5}}. 
\en
Then, performing the integration, we get
\be
F^{(5)}_{\epsilon}=-\frac{\zeta (5)}{2^{5/2}\epsilon^{3/2}}
\frac{f^{2}(r_{0})r^{11/2}_{0}}{\beta^{5}}
\en
The entropy corresponding to this term can be finally obtained
\be
S^{(5)}_{\epsilon}=\beta^{2}\frac{\partial
  F^{(5)}_{\epsilon}}{\partial \beta }= 
\frac{5\zeta (5)}{2^{5/2}\epsilon^{3/2}}
\frac{f^{2}(r_{0})r^{11/2}_{0}}{\beta^{4}}.
\en
For the black hole with the metric (\ref{eq:cinco}) the Hawking temperature 
is $\cite{bhole}$   
\be
\beta_{H}=2\pi r_{0}f^{1/2}(r_{0}).
\en
Introducing this result into the entropy equation, we arrive at the expression
\be
S^{(5)}_{\epsilon}=\frac{5 \zeta
  (5)}{2^{5/2}16\pi^{4}}\frac{r^{3/2}_{0}}{\epsilon^{3/2}}. 
\en
Using again the invariant distance, defined as
\be
\int ds=\int^{r_{0}+\epsilon}_{r_{0}} dr \frac{f^{1/6}(r)}{\sqrt{1-
r^{2}_{0}/r^{2}}}=\sqrt{8f^{1/3}(r_{0})r_{0}\epsilon} ,
\en
the final expression for the entropy is
\be
S^{(5)}_{\epsilon}=\frac{A}{C_{N}8^{3/2}\epsilon
^{3/2}r^{3/2}_{0}f^{1/2}(r_{0})}=\frac{A}{C_{N}(\int  ds)^{3}}, 
\en
where $A=2\pi^{2} r^{3}_{0}f^{1/2}(r_{0})$ and
$C_{N}=\frac{8\pi^{6}}{5 \zeta (5)}$. Therefore the final entropy,
corresponding to the first term of the free energy, is given by 
\be
S_{bare}+S^{(5)}_{\epsilon}=(\frac{1}{4G_{N}}+\frac{1}{(C_{N}\int ds)^{3}})A =
\frac{A}{4G_{R}}.
\en
Let us now calculate the contribution to the free energy, near the
horizon, from the last two terms of the $l$ integration, that is, 
\be
F^{(5)}_{2+3}=-\frac{r^{3}_{0}f^{3/2}(1)}{16\pi}\int dE
\frac{E^{3}}{exp(\beta)-1} 
\int^{\bar{L}^{2}}_{(1+\bar{\epsilon})^{2}} dy (y-1)^{-2}
\en
Using (\ref{eq:energia}) and making the integration, we get  
\be
F^{(5)}_{2+3}=-\frac{\pi^{3}}{480\epsilon}
\frac{r^{4}_{0}f^{3/2}(r_{0})}{\beta^{4}}.
\en
Then, the entropy related to these two terms is
\be
S^{(5)}_{2+3}=\frac{r_{0}}{960\epsilon}=
\frac{r^{2}_{0}f^{1/3}}{960\epsilon r_{0}f^{1/3}}=
\frac{A^{2/3}}{D_{N}(\int ds )^{2}}
\en
where $D_{N}=(2\pi^{2})^{2/3}120$.

To summarize, we have computed the expression for the entropy of scalar fields propagating outside of black holes in four and five dimensions. We have found that the leading divergent term of the entropy is proportional to the area of the horizon, in each case. In the $5$-dimensional case however a subleading divergences arises, proportional to the $\frac{2}{3}$th power of the "area", which in the $4+1$ dimensional case has one(lenght) dimension less than the area itself. Being a subleading divergence with one power less in $(\int ds)^{-1}$, it should not spoil the area law in case we could hide such a term in a coupling redefinition. This is however not the case here. Therefore we are either finding a counter example to the area law in $5$ dimensional gravity, or the brick wall method fails to give the exact answer.  
Some results in the same direction were obtained in \cite{shira},
where subleading corrections to the entropy have also been computed. See also\cite{wintan},
\cite{bin},\cite{wang}.

ACKNOWLEDGMENT:
This work was partially supported
by Funda\c c\~ao de Amparo a Pesquisa do Estado de S\~ao Paulo (FAPESP),
Conselho Nacional de Desenvolvimento  
Cient\'{\i}fico e Tecnol\'{o}gico (CNPQ) and CAPES (Ministry of Education,
Brazil).


\end{document}